\documentclass{PoS}
\usepackage[utf8]{inputenc}
\usepackage{xspace}
\usepackage{verbatim}
\usepackage{subcaption}
\usepackage{caption}
\def\etacar{$\eta$~Carinae\xspace}
\def\gr{$\gamma$-ray\xspace}
\def\grs{$\gamma$-rays\xspace}

\title{Gamma-rays and positrons from Colliding Wind Binaries}

\ShortTitle{Gamma-rays and positrons from CWBs}

\author{\speaker{Matteo Balbo}\\
        Department of Astronomy, University of Geneva, Chemin d’Ecogia 16, 1290 Versoix, Switzerland\\
        E-mail: \email{Matteo.Balbo@unige.ch}}

\author{Roland Walter\\
        Department of Astronomy, University of Geneva, Chemin d’Ecogia 16, 1290 Versoix, Switzerland\\
        E-mail: \email{Roland.Walter@unige.ch}}

\abstract{The \etacar binary system is the first \gr binary ever observed which does not contain a compact object. It is a perfect laboratory to study particle acceleration and \gr emission. Indeed the dense wind of the primary star shocks against the fast light wind coming from the companion star, creating the conditions to accelerate particles up to relativistic energies via Fermi mechanisms. These relativistic particles subsequently dissipate non-thermal radiation. Fermi-LAT and H.E.S.S. detection of \etacar confirm such hypotheses.

Hydrodynamic simulations provide a convincing match to the observations if a few percent of the wind mechanical energy dissipated in the shock goes into particle acceleration. The intrinsic $\pi^0$ decay spectrum is a complex convolution of the maximum energy, luminosity, particle drift and obscuration. Accelerated particles cool down mainly via inverse-Compton, synchrotron radiation, and photo-pion production. High-energy \grs interact also with the pool of anisotropic UV photons emitted by both luminous stars, creating $e^\pm$ pairs and strongly modifying the observed spectrum. Quick variations of the optical depth are expected along the orbit, due to changes in shape, position, and gas density of the shocked region. Various CTA simulations confirm that flux variabilities down to few days timescale could be detected above 30 GeV. These variations could disentangle the intrinsic particle spectral cut off from that related to $\gamma$-$\gamma$ opacity and determine the flux of relativistic protons and positrons injected in the interstellar medium, the geometry of the colliding wind region and the magnetic field configuration, as well as the geometrical orientation of the binary system. CTA will also enlighten the nature of the high-energy component, the mechanisms and the percentage of kinetic energy channelled into particle acceleration.}

\FullConference{High Energy Phenomena in Relativistic Outflows VII - HEPRO VII\\
		9-12 July 2019\\
		Facultat de Física, Universitat de Barcelona, Spain}

\begin{document}

\section{Colliding wind binaries}\index{CWB}

A colliding wind binary (CWB) is a binary star system consisting of two non-compact objects, likely OB-type or Wolf–Rayet~(WR) stars. WRs represent the final stage of the evolution of the most massive stars, before ending their lives as supernovae. Such stars possess the highest known mass loss rates ($10^{-8}\sim10^{-3}~\mathrm{M_\odot/yr}$), and their winds can reach terminal velocities ($v_\infty$) up to few thousands km/s \cite{1987ARA&A..25..113A,1978ARA&A..16..371C}. The kinetic energy of the supersonic wind $L_w = \dot{M}v^2_\infty /2$ rarely exceeds 1\% of the bolometric radiative energy output, which typically is of the order of $\sim 10^{38}~\mathrm{erg/s}$. The radiation pressure drives those powerful winds and creates two strong shock fronts, where the compressed plasma heats up to $\sim10^7~\mathrm{K}$ and emits principally in \mbox{X-rays} \cite{2007A&ARv..14..171D}. In addition to the thermal X-ray emitted in the strong shocks where winds collide, in these regions particles can potentially reach high energies (HE) via diffusive shock acceleration (DSA) \cite{1993ApJ...402..271E}, and they can be exposed to high magnetic field from the hot stars. The existence of the star magnetic field is confirmed by the detected radio synchrotron emission \cite{1986ApJ...303..239A}. Such relativistic particles can also interact with the large pool of soft photons (from IR to UV), which are up-scattered via inverse-Compton (IC) to the GeV energy band, cooling the energetic particles. Other possible cooling mechanisms are bremsstrahlung, photo-pion production or synchrotron radiation \cite{2013A&A...558A..28D}.

\section{\etacar: a PeVatron laboratory}

The \etacar\footnote{The first (optical) observation of the system dates back to 1603, and its original name was $\eta$ Argus.} binary system raised quite some interest in the recent years, after becoming the first binary system ever detected emitting HE \grs, without hosting a compact source. Its main star is supposed to be the most massive and most luminous star of our Galaxy, and is thought to be a Luminous Blue Variable (LBV). Its initial mass is expected to be $\gtrsim 90~M_{\odot}$ \cite{2001ApJ...553..837H}, possibly even up to $\sim 200~\mathrm{M_\odot}$.  For many centuries it was thought to be a single star, and its optical brightness was showing several fluctuations, with a gradually increase up to $18^{th}$ century. Starting from this century, some recorded data are available to trace a light curve (LC) \cite{1999PASP..111.1124H}. Nowadays such fluctuations are generally associated to pre-eruption shifts in apparent temperature of the star's wind, rather than an intrinsic change in luminosity \cite{1997ARA&A..35....1D}. Observations by Sir John Herschel \cite{2011MNRAS.415.2009S}, confirmed that \etacar underwent the so-called \textit{Great Eruption} between 1837 and 1858, increasing its luminosity by more than a factor ten\footnote{The Great Eruption was almost as bright as a supernova explosion, but contrarily to such type of event, the binary survived. This is why sometimes \etacar is also nicknamed as "\textit{supernova impostor}".}, reaching a total luminosity of the order of $10^{7.3}~\mathrm{L_\odot}$. During this $\sim20$ year eruption \etacar ejected an amount of mass of about $10 \sim 40~M_\odot$ \cite{2010MNRAS.401L..48G} at an average speed of $\sim 650$ km s$^{-1}$ \cite{2003AJ....125.1458S}. The released kinetic energy of the ejected material was $\sim10^{49.7}~\mathrm{ergs}$, similar to its bolometric luminosity. Such great eruption is at the origin of the \textit{Homunculus Nebula}, consisting of two pronounced lobes and a large thin equatorial disk.

A second eruption occurred also between 1887 and 1895, but the released amount of material ($\sim 0.2~\mathrm{M_\odot}$) and its kinetic energy this time was smaller ($10^{46.9}$), while its bolometric luminosity was $10^{48.6}$. Thus \etacar ejected material in the polar and equatorial directions during both eruptions, but with different energies and very likely different physical causes.

We had to wait until the last decades of the century to discover the existence of the companion star, and consequently the binarity of the system and its periodicity of 5.52 year \cite{1996ApJ...460L..49D}. The companion star remains unobserved up to now at any wavelength, but it is supposed to be an O-type or WR. Alternative scenario invoked at the origin of the periodicity were pulsation or rotation of the star, thermal/rotational recovery cycle \cite{2005ASPC..332..101D}, but the existence of a companion star is nowadays well established.

The LBV is accelerating a very dense wind with a mass loss rate of \mbox{$\sim 8.5\times 10^{-4}~\mathrm{M_{\odot}~yr^{-1}}$} and a slow terminal velocity of $\sim 420~\mathrm{km~s^{-1}}$ \cite{2012MNRAS.423.1623G}. The WR\index{WR} also emits a powerful wind, less dense but faster, with mass loss rates of $\simeq 10^{-5}~\mathrm{M_{\odot}~yr^{-1}}$ and velocities of $\sim 3000~\mathrm{km~s^{-1}}$ \cite{2002A&A...383..636P}. X-ray observations of the system have been regularly performed since 1997, with the Rossi \mbox{X-ray} Timing Explorer (RXTE) \cite{1993A&AS...97..355B} with nearly weekly intervals (up to daily frequencies during important orbital phase events), which have refined the current\footnote{The orbital period at the epoch of the Great Eruption was estimated to be around $\sim$~5.1 yr.} periodicity to $2024 \pm 2$~days \cite{2008MNRAS.384.1649D}. The modulation detected in the X-ray LC indicates that the two stars are on a very eccentric orbit with $e \simeq 0.9 \sim 0.95$ \cite{2008MNRAS.388L..39O}.

\section{Orbital variability: from X-ray to GeV}

\begin{figure}[!t]
  \begin{center}
    \includegraphics[width=0.8\textwidth]{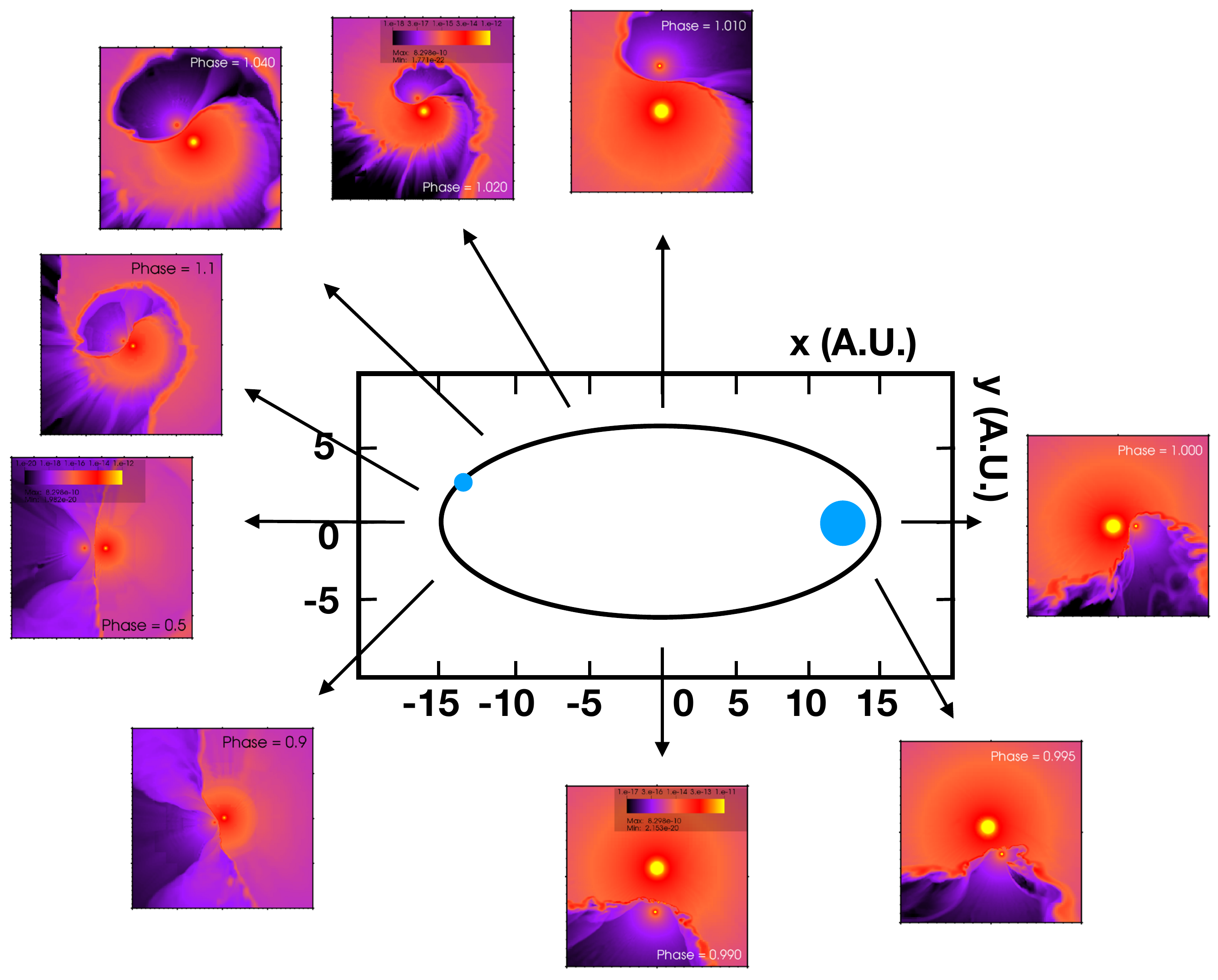}
  \end{center}
  \caption{Different hydro-dynamical simulations of \cite{2011ApJ...726..105P} for different orbital phases of $\eta$~Carinae.}
  \label{fig:orbit_simulation}
\end{figure}

The high estimated eccentricity of the \etacar orbit implies that the relative separation of the two stars varies by a factor $10 \sim 20$ during a full period. At periastron, the two objects pass within a few AU of each other, a distance just a few times larger than the size of the primary star. In the colliding wind region (CWR) the accelerated particles encounter conditions that vary along the orbit: particle density, pressure, gas temperature, local magnetic field, separation between the two stars, etc. (see Fig.~\ref{fig:orbit_simulation}), so an orbital dependency of the $\gamma$-ray emission can also be expected.

The collision of the two stellar winds well explains the X-ray emission modulated by the orbit. The shocked gas can cool by expansion flowing away from the stagnation point along the contact discontinuity, or by emission of radiation. During different orbital phases, the cooling can be dominated by expansion or by radiation. In the latter case, the shock structure becomes unstable \cite{1992ApJ...386..265S} and this is expected to happen when \etacar is close to periastron \cite{2011ApJ...726..105P}. Another aspect in favour of the CWR disruption scenario is the high orbital velocity with which the companion approaches the main star during periastron ($300 \sim 400$~km/s) combined with the extremely small distance separation between the two stars.


Some cycle-to-cycle variations are present around different periastron passages \cite{2015arXiv150707961C}. This is particularly evident in the 2009 periastron event, when the recovery from the X-ray minimum lasted only 1.5 months (nearly half of the usual duration), and in the 2014.5 periastron passage, when the X-ray maximum reached a significantly higher flux. Such variation could be interpreted as small scale variation of the geometry of the shock winds, or with a lower/higher clumpiness degree of the two interacting winds. Cooling processes reduce the downwind gas pressure, shrinking the shocked gas into a thinner layer. The plasma flowing over this thin layer can encounter transverse acceleration instabilities \cite{1993A&A...267..155D} and/or thin-layer instabilities \cite{2000Ap&SS.274..189F}. The observed long-term X-ray modulation, instead, is eventually associated with the time evolution of the ejecta.

The hard X-ray emission detected by INTEGRAL \cite{2008A&A...477L..29L} and Suzaku \cite{2008MNRAS.388L..39O}, with an average luminosity $(4 \sim 7)\times10^{33}~\mathrm{erg~s^{-1}}$, suggested the presence of relativistic particles in the system. With Fermi-LAT we detected two components in the spectrum around periastron passage 2009 \cite{2017A&A...603A.111B}, associating the low energy (LE) component to the IC emission, which cuts off at around few $\sim$ GeV, and well correlates with the orbital motion. Instead, the most energetic component is likely attributed to $\pi^0$ decay (resulting from $pp$ collisions) and can be strongly modulated by $\gamma$-$\gamma$ absorption. Its emission vary by a factor $3 \sim4$ from periastron to apastron, but the limited statistic of the data does not show any evident correlation with the orbital phase. Cherenkov observations also confirmed the detection of \etacar in the highest energy part of the spectrum \cite{2017ICRC...35..717L}, and implies a sudden drop around $\gtrsim 1$ TeV. This could be interpreted either as a cut off in the intrinsic accelerated particle distribution or due to severe $\gamma$-$\gamma$ absorption. 

Using the Hydro-dynamical simulation results which model the interacting stellar winds at each orbital phase \cite{2011ApJ...726..105P}, we have calculated the shock velocities, the mechanical power, and the maximum energy that particles could reach in each adaptive cell of the simulation, balancing the different radiative cooling times (synchrotron, IC, Bremsstrahlung, etc.) with the characteristic acceleration time of the DSA.

\begin{figure}[t]
    \centering
    \begin{minipage}{0.5\textwidth}
        \centering
        \includegraphics[width=\textwidth]{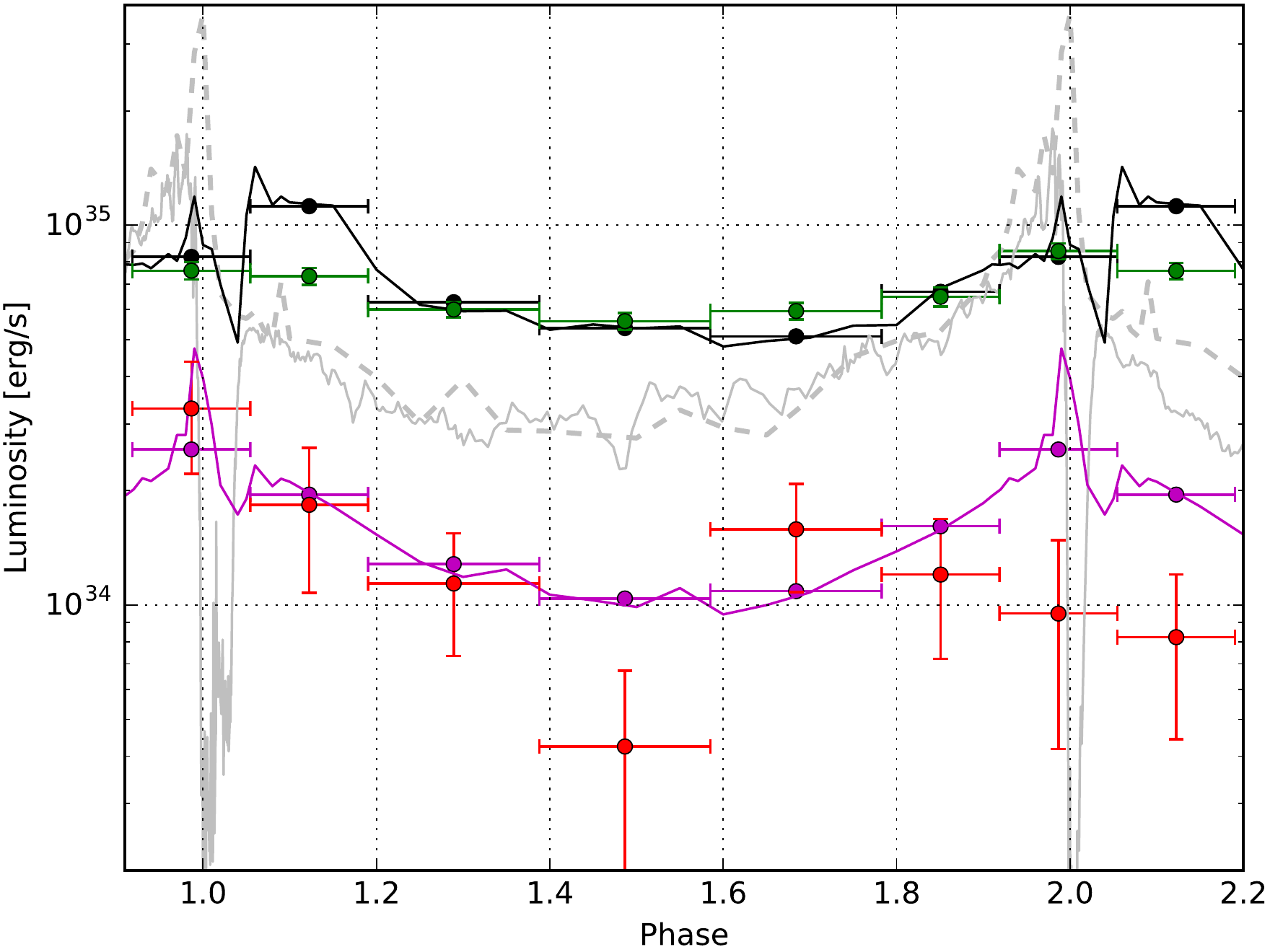}
        \subcaption{}
        \label{fig:LC}
    \end{minipage}%
    \begin{minipage}{0.5\textwidth}
        \centering
        \includegraphics[width=\textwidth]{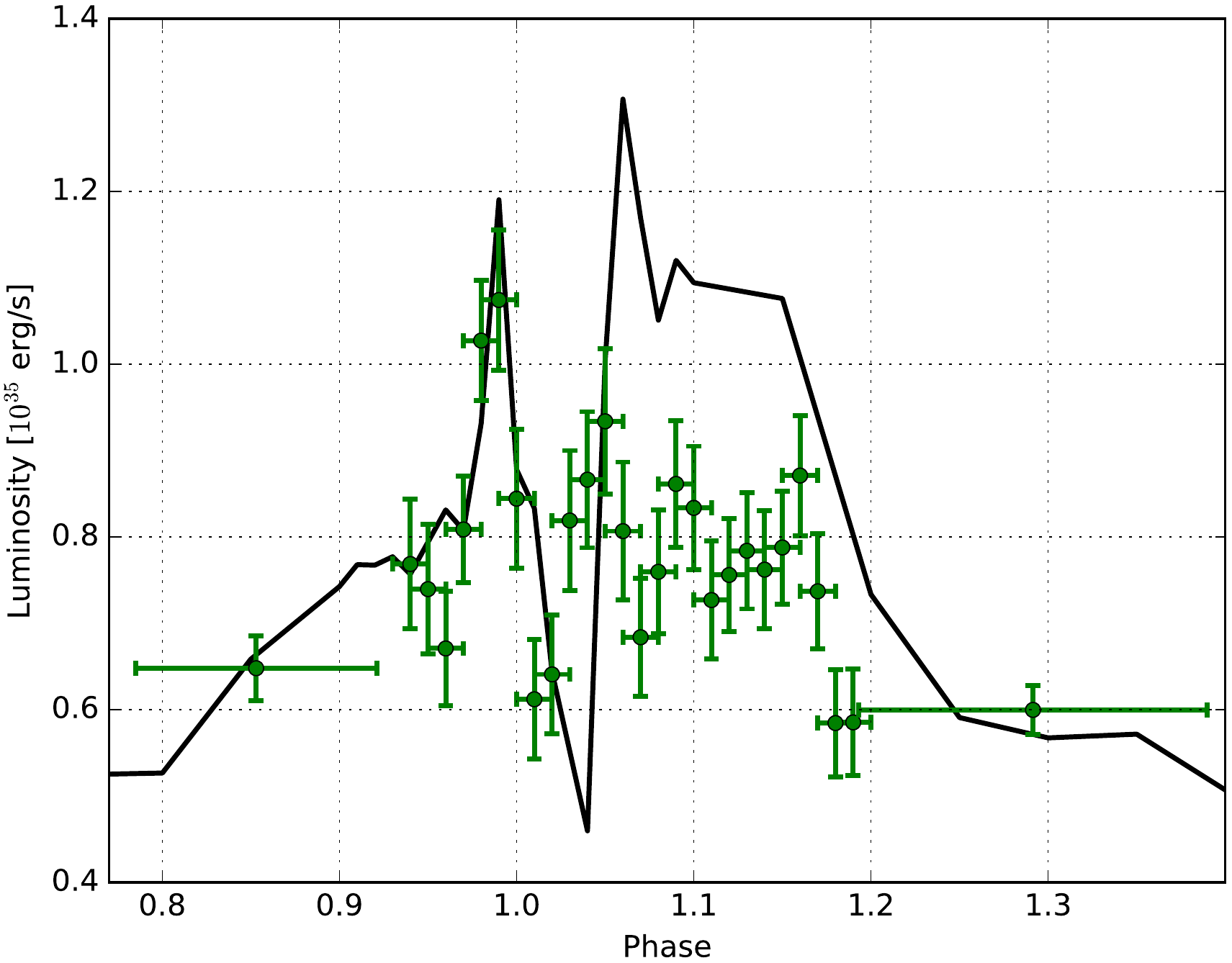}
        \subcaption{}
        \label{fig:LC_zoom}
    \end{minipage}%
\caption{\textbf{(a)} Simulated and observed X-ray and \gr LCs of \etacar. The black and purple lines and bins show the predicted IC and $\pi^0$ decay LCs. The green and red points show the observed Fermi-LAT LC at low (0.3-10 GeV) and high (10-300 GeV) energies. The dim gray lines show the observed (continuous) and predicted (dash, without obscuration) thermal X-ray LCs. Errorbars are $1\sigma$.
\textbf{(b)} A merged Fermi-LAT analysis (0.3-10 GeV) of the two periastrons for narrow time bins. The two broad bins and the black curve are the same as in Fig.~\ref{fig:LC}.}
\label{fig:LC_generic}
\end{figure}

Most of the shock power is released on both sides of the wind collision zone and in the cells downstream the CWR \cite{2006ApJ...644.1118R}. The increasing shock area compensates for the loss of the released energy density up to a relatively large distance from the centre of mass, explaining why the \mbox{X-ray} luminosity at apastron is still about a third of the peak emission at periastron. Computing the emissivity of the resulting electron spectra, we could reproduce the LE (0.3-10 GeV) \gr LC of \etacar very well (see Fig.~\ref{fig:LC}), with the only overestimation of a time interval after the periastron, which suggests that the shock front can in reality be more unstable. In particular, the "reverse bubble cavity" (see Fig.~\ref{fig:orbit_simulation}) that is created by the close passage of the companion with its very fast wind could in reality collapse. The relative importance of this secondary peak depends on the magnetic field geometry, radiation transfer, obscuration and other details of the hydrodynamics, and its spectral shape could in reality be different. The fact that the LE \gr LC showed a similar behaviour during both periastrons, allowed us to perform a stacked analysis below $\lesssim 10~\mathrm{GeV}$, decreasing the temporal resolution down to $\sim 40$ days (see Fig.~\ref{fig:LC_zoom}). 

\begin{figure}[!t]
  \begin{center}
    \includegraphics[width=0.8\textwidth]{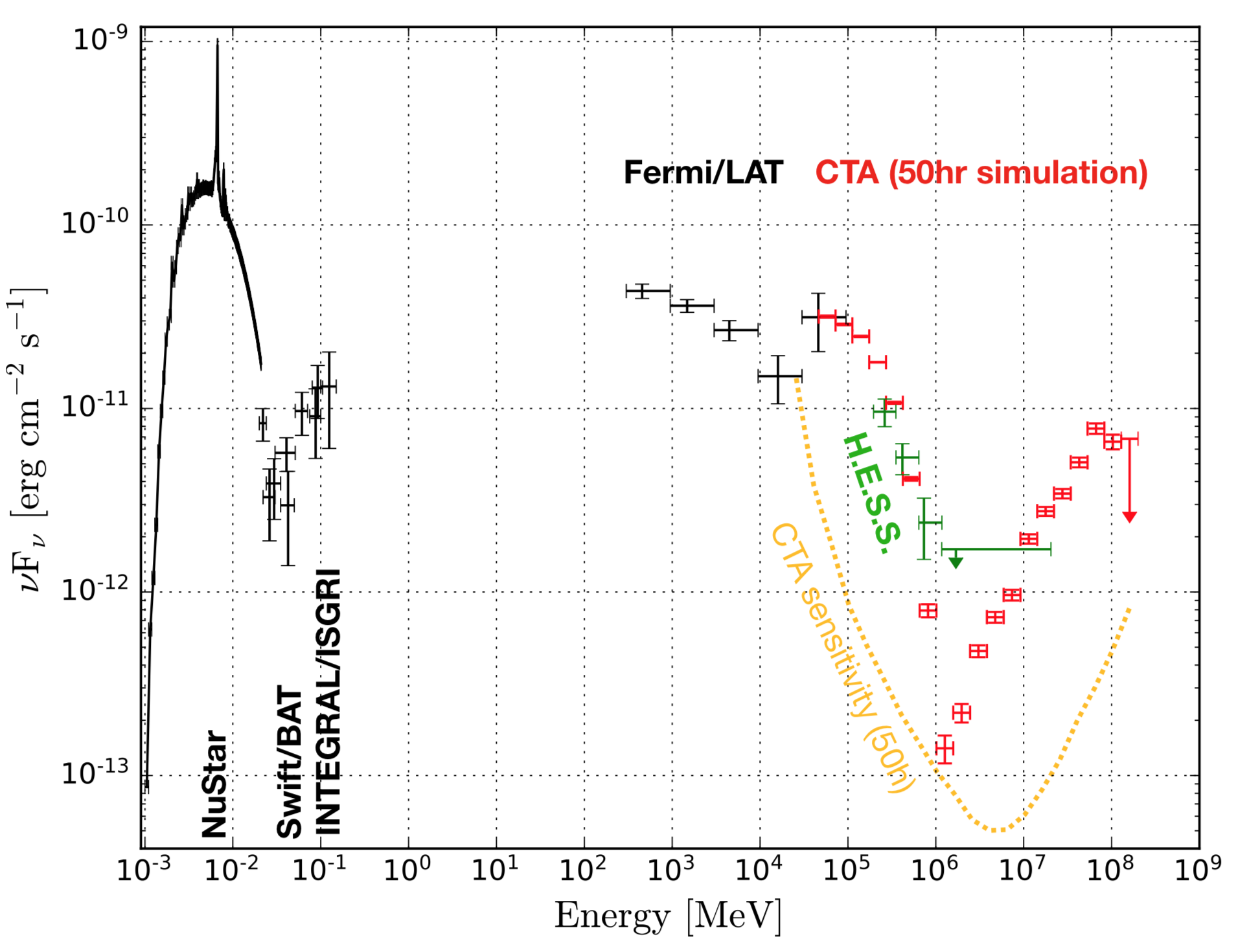}
  \end{center}
  \caption{Spectral energy distribution of \etacar from 1 keV to 100 TeV. The data are from NuStar \cite{2018A&A...610A..37P}, Swift, INTEGRAL, Fermi-LAT and H.E.S.S. and obtained close to periastron. The red points show the results of a simulation of what could be detected by CTA (at periastron) assuming that the emission is dominated by $\pi^0$ decay modified by $\gamma$-$\gamma$ absorption in the strong ultraviolet photon field.}
  \label{fig:etacar_SED}
\end{figure}

The mechanical luminosity available in the shock increases towards periastron, the same trend is followed by the thermal emission and by the LE \grs, almost doubling in the phase range $\approx 1.05 - 1.15$. The latter peak corresponds to the "reverse bubble", which effectively doubled the shock front area during about a tenth of the orbit \cite{2011ApJ...726..105P}. The mechanical luminosity shows also a local minimum between phases 1.0 and 1.05, when the central part of the CWR is disrupted, similarly to the X-ray emission. If the final spectra of accelerated electrons, measured on all cells on the side of the primary star, overlap significantly with the one from the secondary star side, and have a similar luminosity, the resulting \gr spectra due to IC cooling are not capable to explain any prominent feature in the SED \cite{2017A&A...603A.111B}. So in this case leptons can not explain the VHE \gr emission observed during 2009 periastron passage. Furthermore, their contribution seems to cut off at around few $\sim\mathrm{GeV}$ (see Fig.~\ref{fig:etacar_SED}).

The hard component above $\gtrsim 10~\mathrm{GeV}$ could be explained uniquely with an hadronic contribution, and its emission takes place close to the center. Simulations results suggest a surface magnetic field in the range 0.4 - 1 kG, which in turn can be reduced by some order of magnitude if strong amplification is assumed at the shock \cite{2012MNRAS.423.1562F}. Particle acceleration in turbulent reconnection at the current sheets is not excluded \cite{2011ApJ...735..102K} and is favoured by small-scale turbulent motions \cite{2009ApJ...700...63K}. Assuming that $\sim 2.4\%$ of the mechanical energy goes into proton acceleration, \grs emitted by $\pi^0$ decay could well reproduce the variability pattern from periastron 2009 toward apastron 2012. At periastron, protons could reach energy close to $10^{15}~\mathrm{eV}$ (i.e. close to the knee of the cosmic-ray spectrum). Differently from SNRs that can efficiently accelerate particles only for a limited period of their evolution (few $\sim 10^3$ years), CWBs can accelerate particles for a much longer period of their life, providing even up to $10^{48 \sim 49}~\mathrm{erg}$ of cosmic-ray energy.

Despite the variability from periastron 2009 to apastron 2012 was well predicted by the simulation, in the following 2014.5 periastron passage the hard component did not show up again, remaining at a flux level compatible with the apastron. The intrinsic $\pi^0$ decay spectrum is a complex convolution of the maximum energy, luminosity, particle drift and obscuration expected in every simulation cell. The scarce statistic of events at HE does not allow to reduce the temporal binning of the LC much below several months.


The fraction of the shock mechanical luminosity accelerating electrons appears to be slightly smaller than the one that accelerates protons. These results contrast with the efficiencies derived from the latest PIC simulations \cite{2015PhRvL.114h5003P}, involving low magnetic fields, radiation and particle densities and favouring hadronic acceleration in the context of SNR. Purely hadronic acceleration has been proposed by \cite{2015MNRAS.449L.132O} to explain the GeV spectrum of \etacar. In that case the two spectral components are related to the different hadron interaction times observed on the two sides of the wind separation surface, largely because of the contrast in density and magnetic field. Similarly to what happens to electrons, in our simulations this effect is smoothed by the many zones of the model, each characterized by different conditions. Even if the shock on the companion side does contribute more at high energies, the resulting $\pi^0$ decay spectrum does not feature two distinct components. An instrument sensitive in the LE \gr band (1-100~MeV), would easily discriminate between the lepto-hadronic and purely hadronic models. Indeed, the IC \gr emission of the primary electron component would be much stronger than the one predicted by secondary leptons produced in the purely hadronic acceleration scenario.

\section{Opacity study for high-energy $\gamma$-rays and CTA prospects}
	\label{sec:eta_car_opacity}

The multiwavelength spectrum of \etacar is quite well understood, from radio to X-ray. Instead, for energy above the MeV domain the correct interpretation of the spectrum is still debated, mostly due to the very low statistic of the available data. The clear presence of an hard component around periastron 2009 above $\gtrsim 10$~GeV can hardly be explained invoking a leptonic scenario \cite{2017A&A...603A.111B}. The observed feature around $\sim 10$~GeV would require an ad hoc distribution of X-ray photons to play a significant role with the keV-GeV photon absorption \cite{2012A&A...544A..98R}. In fact, the opacities measured for \mbox{X-ray} and \gr photons in our simulation \cite{2017A&A...603A.111B}, along different possible lines of sight, yield always a negligible value of $\tau<10^{-2}$ around periastron, and many orders of magnitude smaller during all other phases.

Photo-pion production could potentially derive from $p \gamma$ or \textit{pp} interactions. The former presents a cross section two orders of magnitude smaller than the latter one, and its peak occurs when $E_\gamma \cdot E_p \simeq 3.5 \times 10^{17}~\mathrm{eV^2}$. As the maximum of the thermal soft photons is emitted in the UV energy band, the required proton energy would result to be extremely high $E_p \simeq 10\sim100$~PeV, thus making such process less likely to happen. Less energetic protons instead are required if the interactions happen with more energetic \grs (e.g. those UV that are up-scattered via IC scattering off relativistic electrons).

Above several hundreds of GeV, $\gamma$-$\gamma$ absorption and subsequent $e^\pm$ pairs production could arise from eV-TeV photon interactions, making the photon field opaque close to $\sim \mathrm{TeVs}$. The region is indeed constantly replenished by thermal UV photons from the massive star, whose density reduces for higher distances $\propto 1/r^2$. The obscuration is likely maximized at periastron, when the ultraviolet photon field is particularly dense and the intrinsic cut off energy of the particle spectra is the highest. The probability of interaction for the two photons depends on the density of the UV and \grs, on the scattering angle between the two interacting photons, and on the orientation of the system with respect to the observer. The location where relativistic \grs are produced with respect to the main star and the observer changes as a direct consequence of the orbital motion. So HE photons travelling toward us can interact along their path with UV photons at angles which vary according to the different orbital phases. For a given UV photon of energy $\sim k_B T$ and a \gr of energy $E_\gamma$, the $\gamma$-$\gamma$ absorption can be expressed as a function of $\xi = E_\gamma \cdot k_B T / (m_e c^2)^2$. If collisions occur mostly head-on, the peak of the photon-photon absorption is located around $\xi \simeq 1.4$, whereas in case of tail-in collisions it moves towards $\xi \simeq 30$ \cite{2018MNRAS.474.1436V}. 

\begin{figure}[!t]
  \begin{center}
    \includegraphics[width=0.8\textwidth]{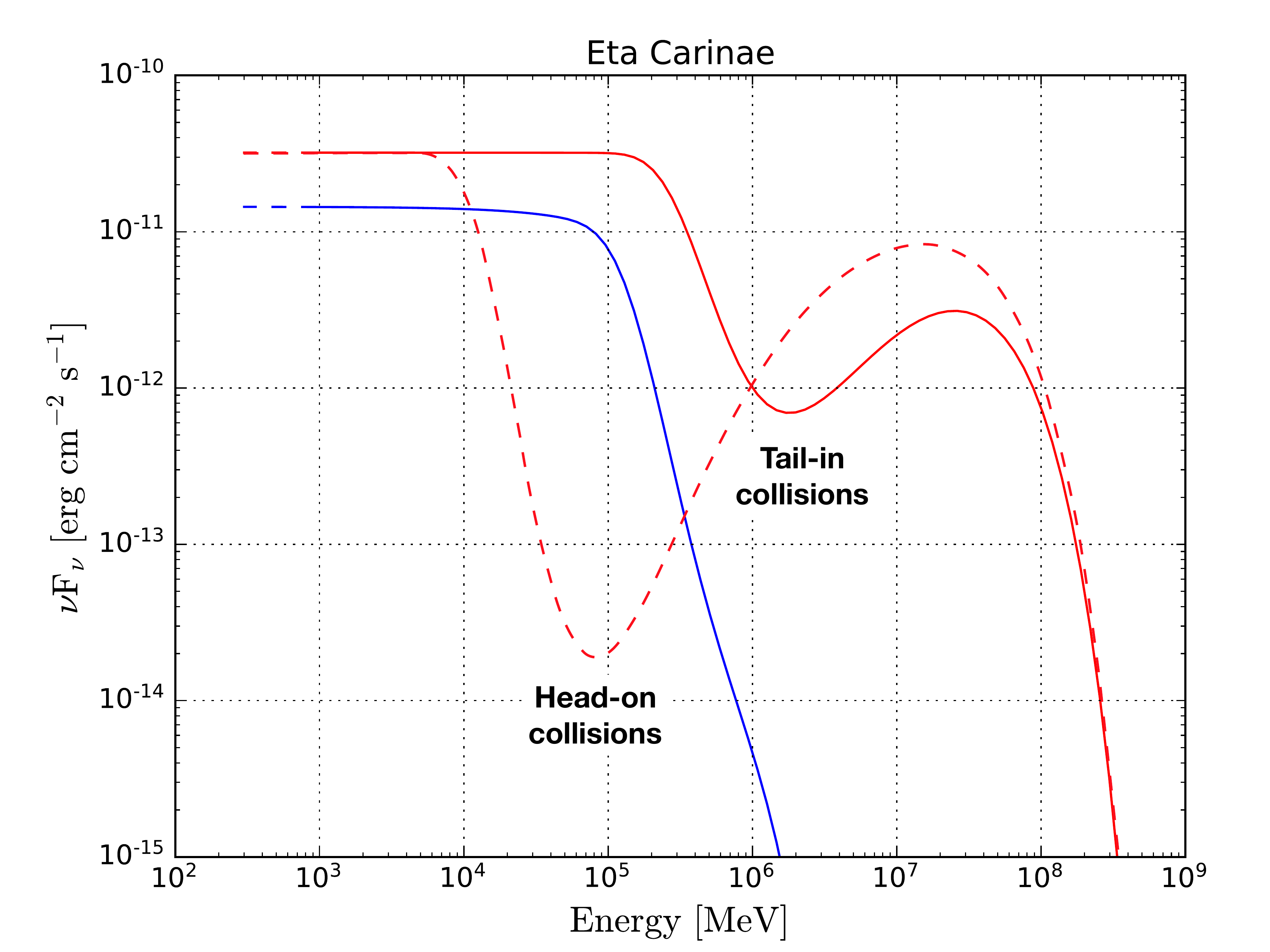}
  \end{center}
  \caption{Possible hadronic \gr emission from \etacar, convoluted with the expected $\gamma$-$\gamma$ absorption and the different particle energy cut off, at around periastron (red) and apastron (blue). The dashed line shows the theoretical variability expected at around periastron \cite[in preparation]{2019...Balbo...paper}.}
  \label{fig:etacar_SED_abs}
\end{figure}

Using the most likely orientation of the binary system \cite{2012MNRAS.420.2064M}, the strongest absorption variability is supposed to occur around periastron. Approaching periastron, indeed,  the companion star lies between the main star and the observer. Once periastron is passed, the companion star moves further away from the observer and the main star. The two stars are roughly aligned at phases 0.9 and 1.1. The HE photons produced by the colliding winds will propagate towards the observer initially interacting with the strong ultraviolet field, mostly with “tail-in” collisions. Once periastron is passed, the collisions will mainly occur “head-on” from the shocked region up to the main star, and “tail-in” when the photons will propagate further. The final $\gamma$-$\gamma$ absorption varies by nearly a factor 40 in intensity when we pass from “head-on” to “tail-in” collisions [34]. Furthermore, the energy of the maximum absorption varies by a factor > 20. Assuming a black-body emission for the main star with a temperature of 37000~K, the peak of the absorption spans from few tens up to several hundreds GeV. The combination of all these effects will result in a strong modulation of the $\gamma$-$\gamma$ absorption around periastron, as shown in Fig.~\ref{fig:etacar_SED_abs}. A modification of the black-body emission by absorption in the wind will move the peak of the $\gamma$-$\gamma$ absorption toward even higher energies.

A large combination of parameters is possible, but quality of current data (Fermi-LAT and H.E.S.S.) do not allow to refine any model parameter. The more performing CTA sensitivity (many orders of magnitude) with respect to Fermi-LAT, in particular for very short observations \cite{2013APh....43..348F}, will allow to decrease enormously the temporal binning of the LC, offering an unprecedented level of details and S/N for variability studies and spectral analyses.

In Fig.~\ref{fig:etacar_SED} a 50 hours CTA simulation of a an arbitrary spectrum around periastron is reported. It has been obtained using the CTOOLS and \texttt{Prod3b-v2} IRF of the South array\footnote{I used the \texttt{South\_z40\_50h} IRF.}. Provided that the loosely constrained optical depth is in a reasonable range, the $\pi^0$ decay continuum could optimistically be detected even with very short exposure of $\lesssim$ 1 hour. 

A detailed study of the \etacar LC variability and the VHE cut off modulation will allow to probe the geometry of the system, the magnetic field configuration, the location where protons are accelerated and their maximum energy, as well as the distribution of the soft photon in a very complex environment \cite[submitted]{2019...Masha...paper}. Finally, the study of \etacar provides us with a remarkable opportunity to study close-up one of the most massive stars known during the final stage of its evolution, as well as the strong shock wind interactions with its companion. 



\bibliography{Bibliography}

\providecommand{\href}[2]{#2}\begingroup\raggedright\begin{thebibliography}{10}

\bibitem{1987ARA&A..25..113A}
D.~C. {Abbott} and P.~S. {Conti}, \emph{{Wolf-rayet stars.}},
  \href{https://doi.org/10.1146/annurev.aa.25.090187.000553}{\emph{Annual
  Review of Astronomy and Astrophysics} {\bfseries 25} (1987) 113}.

\bibitem{1978ARA&A..16..371C}
P.~S. {Conti}, \emph{{Mass loss in early-type stars.}},
  \href{https://doi.org/10.1146/annurev.aa.16.090178.002103}{\emph{Annual
  Review of Astronomy and Astrophysics} {\bfseries 16} (1978) 371}.

\bibitem{2007A&ARv..14..171D}
M.~{De Becker}, \emph{{Non-thermal emission processes in massive binaries}},
  \href{https://doi.org/10.1007/s00159-007-0005-2}{\emph{Astronomy and
  Astrophysics Review} {\bfseries 14} (2007) 171}
  [\href{https://arxiv.org/abs/0709.4220}{{\ttfamily 0709.4220}}].

\bibitem{1993ApJ...402..271E}
D.~{Eichler} and V.~{Usov}, \emph{{Particle Acceleration and Nonthermal Radio
  Emission in Binaries of Early-Type Stars}},
  \href{https://doi.org/10.1086/172130}{\emph{\apj} {\bfseries 402} (1993)
  271}.

\bibitem{1986ApJ...303..239A}
D.~C. {Abbott}, J.~H. {Beiging}, E.~{Churchwell} and A.~V. {Torres},
  \emph{{Radio Emission from Galactic Wolf-Rayet Stars and the Structure of
  Wolf-Rayet Winds}}, \href{https://doi.org/10.1086/164070}{\emph{\apj}
  {\bfseries 303} (1986) 239}.

\bibitem{2013A&A...558A..28D}
M.~{De Becker} and F.~{Raucq}, \emph{{Catalogue of particle-accelerating
  colliding-wind binaries}},
  \href{https://doi.org/10.1051/0004-6361/201322074}{\emph{\aap} {\bfseries
  558} (2013) A28} [\href{https://arxiv.org/abs/1308.3149}{{\ttfamily
  1308.3149}}].

\bibitem{2001ApJ...553..837H}
D.~J. {Hillier}, K.~{Davidson}, K.~{Ishibashi} and T.~{Gull}, \emph{{On the
  Nature of the Central Source in {$\eta$} Carinae}},
  \href{https://doi.org/10.1086/320948}{\emph{\apj} {\bfseries 553} (2001)
  837}.

\bibitem{1999PASP..111.1124H}
R.~M. {Humphreys}, K.~{Davidson} and N.~{Smith}, \emph{{{\ensuremath{\eta}}
  Carinae's Second Eruption and the Light Curves of the {\ensuremath{\eta}}
  Carinae Variables}}, \href{https://doi.org/10.1086/316420}{\emph{Publications
  of the Astronomical Society of the Pacific} {\bfseries 111} (1999) 1124}.

\bibitem{1997ARA&A..35....1D}
K.~{Davidson} and R.~M. {Humphreys}, \emph{{Eta Carinae and Its Environment}},
  \href{https://doi.org/10.1146/annurev.astro.35.1.1}{\emph{Annual Review of
  Astronomy and Astrophysics} {\bfseries 35} (1997) 1}.

\bibitem{2011MNRAS.415.2009S}
N.~{Smith} and D.~J. {Frew}, \emph{{A revised historical light curve of Eta
  Carinae and the timing of close periastron encounters}},
  \href{https://doi.org/10.1111/j.1365-2966.2011.18993.x}{\emph{\mnras}
  {\bfseries 415} (2011) 2009}
  [\href{https://arxiv.org/abs/1010.3719}{{\ttfamily 1010.3719}}].

\bibitem{2010MNRAS.401L..48G}
H.~L. {Gomez}, C.~{Vlahakis}, C.~M. {Stretch}, L.~{Dunne}, S.~A. {Eales},
  A.~{Beelen} et~al., \emph{{Submillimetre variability of Eta Carinae: cool
  dust within the outer ejecta}},
  \href{https://doi.org/10.1111/j.1745-3933.2009.00784.x}{\emph{\mnras}
  {\bfseries 401} (2010) L48}
  [\href{https://arxiv.org/abs/0911.0176}{{\ttfamily 0911.0176}}].

\bibitem{2003AJ....125.1458S}
N.~{Smith}, R.~D. {Gehrz}, P.~M. {Hinz}, W.~F. {Hoffmann}, J.~L. {Hora}, E.~E.
  {Mamajek} et~al., \emph{{Mass and Kinetic Energy of the Homunculus Nebula
  around {$\eta$} Carinae}}, \href{https://doi.org/10.1086/346278}{\emph{\aj}
  {\bfseries 125} (2003) 1458}.

\bibitem{1996ApJ...460L..49D}
A.~{Damineli}, \emph{{The 5.52 Year Cycle of Eta Carinae}},
  \href{https://doi.org/10.1086/309961}{\emph{\apj} {\bfseries 460} (1996)
  L49}.

\bibitem{2005ASPC..332..101D}
K.~{Davidson}, \emph{{The Physical Nature of $\eta$ Carinae}},  in \emph{The
  Fate of the Most Massive Stars}, R.~{Humphreys} and K.~{Stanek}, eds.,
  vol.~332 of \emph{Astronomical Society of the Pacific Conference Series},
  p.~103, Sep, 2005.

\bibitem{2012MNRAS.423.1623G}
J.~H. {Groh}, D.~J. {Hillier}, T.~I. {Madura} and G.~{Weigelt}, \emph{{On the
  influence of the companion star in Eta Carinae: 2D radiative transfer
  modelling of the ultraviolet and optical spectra}},
  \href{https://doi.org/10.1111/j.1365-2966.2012.20984.x}{\emph{\mnras}
  {\bfseries 423} (2012) 1623}
  [\href{https://arxiv.org/abs/1204.1963}{{\ttfamily 1204.1963}}].

\bibitem{2002A&A...383..636P}
J.~M. {Pittard} and M.~F. {Corcoran}, \emph{{In hot pursuit of the hidden
  companion of eta Carinae: An X-ray determination of the wind parameters}},
  \href{https://doi.org/10.1051/0004-6361:20020025}{\emph{\aap} {\bfseries 383}
  (2002) 636} [\href{https://arxiv.org/abs/astro-ph/0201105}{{\ttfamily
  astro-ph/0201105}}].

\bibitem{1993A&AS...97..355B}
H.~V. {Bradt}, R.~E. {Rothschild} and J.~H. {Swank}, \emph{{X-ray timing
  explorer mission}}, {\emph{Astronomy and Astrophysics Supplement Series}
  {\bfseries 97} (1993) 355}.

\bibitem{2008MNRAS.384.1649D}
A.~{Damineli}, D.~J. {Hillier}, M.~F. {Corcoran}, O.~{Stahl}, R.~S.
  {Levenhagen}, N.~V. {Leister} et~al., \emph{{The periodicity of the {$\eta$}
  Carinae events}},
  \href{https://doi.org/10.1111/j.1365-2966.2007.12815.x}{\emph{\mnras}
  {\bfseries 384} (2008) 1649}
  [\href{https://arxiv.org/abs/0711.4250}{{\ttfamily 0711.4250}}].

\bibitem{2008MNRAS.388L..39O}
A.~T. {Okazaki}, S.~P. {Owocki}, C.~M.~P. {Russell} and M.~F. {Corcoran},
  \emph{{Modelling the RXTE light curve of {$\eta$} Carinae from a 3D SPH
  simulation of its binary wind collision}},
  \href{https://doi.org/10.1111/j.1745-3933.2008.00496.x}{\emph{\mnras}
  {\bfseries 388} (2008) L39}
  [\href{https://arxiv.org/abs/0805.1794}{{\ttfamily 0805.1794}}].

\bibitem{2011ApJ...726..105P}
E.~R. {Parkin}, J.~M. {Pittard}, M.~F. {Corcoran} and K.~{Hamaguchi},
  \emph{{Spiraling Out of Control: Three-dimensional Hydrodynamical Modeling of
  the Colliding Winds in {$\eta$} Carinae}},
  \href{https://doi.org/10.1088/0004-637X/726/2/105}{\emph{\apj} {\bfseries
  726} (2011) 105} [\href{https://arxiv.org/abs/1011.0778}{{\ttfamily
  1011.0778}}].

\bibitem{1992ApJ...386..265S}
I.~R. {Stevens}, J.~M. {Blondin} and A.~M.~T. {Pollock}, \emph{{Colliding Winds
  from Early-Type Stars in Binary Systems}},
  \href{https://doi.org/10.1086/171013}{\emph{\apj} {\bfseries 386} (1992)
  265}.

\bibitem{2015arXiv150707961C}
M.~F. {Corcoran}, K.~{Hamaguchi}, J.~K. {Liburd}, D.~{Morris}, T.~R. {Gull},
  T.~I. {Madura} et~al., \emph{{The X-ray Lightcurve of the Supermassive star
  eta Carinae, 1996--2014}}, {\emph{arXiv e-prints} (2015) arXiv:1507.07961}
  [\href{https://arxiv.org/abs/1507.07961}{{\ttfamily 1507.07961}}].

\bibitem{1993A&A...267..155D}
R.~{Dgani}, R.~{Walder} and H.~{Nussbaumer}, \emph{{Stability analysis of
  colliding winds in a double star system.}}, {\emph{\aap} {\bfseries 267}
  (1993) 155}.

\bibitem{2000Ap&SS.274..189F}
D.~{Folini} and R.~{Walder}, \emph{{3D Hydrodynamical Simulations of Colliding
  Wind Binaries: Theory Confronts Observations}},
  \href{https://doi.org/10.1023/A:1026560309386}{\emph{\apss} {\bfseries 274}
  (2000) 189}.

\bibitem{2008A&A...477L..29L}
J.-C. {Leyder}, R.~{Walter} and G.~{Rauw}, \emph{{Hard X-ray emission from
  {$\eta$} Carinae}},
  \href{https://doi.org/10.1051/0004-6361:20078981}{\emph{\aap} {\bfseries 477}
  (2008) L29} [\href{https://arxiv.org/abs/0712.1491}{{\ttfamily 0712.1491}}].

\bibitem{2017A&A...603A.111B}
M.~{Balbo} and R.~{Walter}, \emph{{Fermi acceleration along the orbit of
  {$\eta$} Carinae}},
  \href{https://doi.org/10.1051/0004-6361/201629640}{\emph{\aap} {\bfseries
  603} (2017) A111} [\href{https://arxiv.org/abs/1705.02706}{{\ttfamily
  1705.02706}}].

\bibitem{2017ICRC...35..717L}
E.~{Leser}, S.~{Ohm}, M.~{F{\"u}{\ss}ling}, M.~{de Naurois}, K.~{Egberts},
  P.~{Bordas} et~al., \emph{{First Results of Eta Carinae Observations with
  H.E.S.S. II}}, {\emph{International Cosmic Ray Conference} {\bfseries 301}
  (2017) 717} [\href{https://arxiv.org/abs/1708.01033}{{\ttfamily
  1708.01033}}].

\bibitem{2006ApJ...644.1118R}
A.~{Reimer}, M.~{Pohl} and O.~{Reimer}, \emph{{Nonthermal High-Energy Emission
  from Colliding Winds of Massive Stars}},
  \href{https://doi.org/10.1086/503598}{\emph{\apj} {\bfseries 644} (2006)
  1118} [\href{https://arxiv.org/abs/astro-ph/0510701}{{\ttfamily
  astro-ph/0510701}}].

\bibitem{2018A&A...610A..37P}
C.~{Panagiotou} and R.~{Walter}, \emph{{The environment of the wind-wind
  collision region of {\ensuremath{\eta}} Carinae}},
  \href{https://doi.org/10.1051/0004-6361/201731841}{\emph{\aap} {\bfseries
  610} (2018) A37} [\href{https://arxiv.org/abs/1712.01382}{{\ttfamily
  1712.01382}}].

\bibitem{2012MNRAS.423.1562F}
D.~{Falceta-Gon{\c{c}}alves} and Z.~{Abraham}, \emph{{MHD numerical simulations
  of colliding winds in massive binary systems - I. Thermal versus non-thermal
  radio emission}},
  \href{https://doi.org/10.1111/j.1365-2966.2012.20978.x}{\emph{\mnras}
  {\bfseries 423} (2012) 1562}
  [\href{https://arxiv.org/abs/1203.5093}{{\ttfamily 1203.5093}}].

\bibitem{2011ApJ...735..102K}
G.~{Kowal}, E.~M. {de Gouveia Dal Pino} and A.~{Lazarian},
  \emph{{Magnetohydrodynamic Simulations of Reconnection and Particle
  Acceleration: Three-dimensional Effects}},
  \href{https://doi.org/10.1088/0004-637X/735/2/102}{\emph{\apj} {\bfseries
  735} (2011) 102} [\href{https://arxiv.org/abs/1103.2984}{{\ttfamily
  1103.2984}}].

\bibitem{2009ApJ...700...63K}
G.~{Kowal}, A.~{Lazarian}, E.~T. {Vishniac} and K.~{Otmianowska-Mazur},
  \emph{{Numerical Tests of Fast Reconnection in Weakly Stochastic Magnetic
  Fields}}, \href{https://doi.org/10.1088/0004-637X/700/1/63}{\emph{\apj}
  {\bfseries 700} (2009) 63} [\href{https://arxiv.org/abs/0903.2052}{{\ttfamily
  0903.2052}}].

\bibitem{2015PhRvL.114h5003P}
J.~{Park}, D.~{Caprioli} and A.~{Spitkovsky}, \emph{{Simultaneous Acceleration
  of Protons and Electrons at Nonrelativistic Quasiparallel Collisionless
  Shocks}}, \href{https://doi.org/10.1103/PhysRevLett.114.085003}{\emph{\prl}
  {\bfseries 114} (2015) 085003}
  [\href{https://arxiv.org/abs/1412.0672}{{\ttfamily 1412.0672}}].

\bibitem{2015MNRAS.449L.132O}
S.~{Ohm}, V.~{Zabalza}, J.~A. {Hinton} and E.~R. {Parkin}, \emph{{On the origin
  of {$\gamma$}-ray emission in {$\eta$} Carina}},
  \href{https://doi.org/10.1093/mnrasl/slv032}{\emph{\mnras} {\bfseries 449}
  (2015) L132} [\href{https://arxiv.org/abs/1502.04056}{{\ttfamily
  1502.04056}}].

\bibitem{2012A&A...544A..98R}
K.~{Reitberger}, O.~{Reimer}, A.~{Reimer}, M.~{Werner}, K.~{Egberts} and
  H.~{Takahashi}, \emph{{Gamma-ray follow-up studies on {$\eta$} Carinae}},
  \href{https://doi.org/10.1051/0004-6361/201219249}{\emph{\aap} {\bfseries
  544} (2012) A98} [\href{https://arxiv.org/abs/1203.4939}{{\ttfamily
  1203.4939}}].

\bibitem{2018MNRAS.474.1436V}
G.~{Voisin}, F.~{Mottez} and S.~{Bonazzola}, \emph{{Electron-positron pair
  production by gamma-rays in an anisotropic flux of soft photons, and
  application to pulsar polar caps}},
  \href{https://doi.org/10.1093/mnras/stx2658}{\emph{\mnras} {\bfseries 474}
  (2018) 1436} [\href{https://arxiv.org/abs/1710.04021}{{\ttfamily
  1710.04021}}].

\bibitem{2012MNRAS.420.2064M}
T.~I. {Madura}, T.~R. {Gull}, S.~P. {Owocki}, J.~H. {Groh}, A.~T. {Okazaki} and
  C.~M.~P. {Russell}, \emph{{Constraining the absolute orientation of
  {\ensuremath{\eta}} Carinae's binary orbit: a 3D dynamical model for the
  broad [Fe III] emission}},
  \href{https://doi.org/10.1111/j.1365-2966.2011.20165.x}{\emph{\mnras}
  {\bfseries 420} (2012) 2064}
  [\href{https://arxiv.org/abs/1111.2226}{{\ttfamily 1111.2226}}].

\bibitem{2019...Balbo...paper}
M.~{Balbo} and R.~{Walter}, \emph{{$\gamma$-$\gamma$ opacity study in
  \etacar}}, .

\bibitem{2013APh....43..348F}
S.~{Funk}, J.~A. {Hinton} and {CTA Consortium}, \emph{{Comparison of Fermi-LAT
  and CTA in the region between 10-100 GeV}},
  \href{https://doi.org/10.1016/j.astropartphys.2012.05.018}{\emph{Astroparticle
  Physics} {\bfseries 43} (2013) 348}
  [\href{https://arxiv.org/abs/1205.0832}{{\ttfamily 1205.0832}}].

\bibitem{2019...Masha...paper}
M.~{Chernyakova}, D.~{Malyshev}, A.~{Paizis}, N.~{La Palombara}, M.~{Balbo},
  R.~{Walter} et~al., \emph{{Prospects for observations of non-transient
  $\gamma$-ray binaries with Cherenkov Telescope Array.}}, .

\end{thebibliography}\endgroup

\end{document}